\documentclass[aps,prd,twocolumn,superscriptaddress]{revtex4}
\usepackage{epsfig,epsf}
\usepackage{amsmath}
\usepackage{amsthm}
\usepackage{amsfonts}
\usepackage{amssymb}
\usepackage{dsfont}
\usepackage{multirow}
\usepackage{appendix}
\usepackage{slashed}
\usepackage[active]{srcltx}
\usepackage{psfrag}

\begin{document}

\title{Heavy axial-vector structures $bb\overline{c}\overline{c}$}
\date{\today}
\author{S.~S.~Agaev}
\affiliation{Institute for Physical Problems, Baku State University, Az--1148 Baku,
Azerbaijan}
\author{K.~Azizi}
\affiliation{Department of Physics, University of Tehran, North Karegar Avenue, Tehran
14395-547, Iran}
\affiliation{Department of Physics, Do\v{g}u\c{s} University, Dudullu-\"{U}mraniye, 34775
Istanbul, T\"{u}rkiye}
\author{H.~Sundu}
\affiliation{Department of Physics Engineering, Istanbul Medeniyet University, 34700
Istanbul, T\"{u}rkiye}

\begin{abstract}
The fully heavy axial-vector diquark-antidiquark structures $bb\overline{c}%
\overline{c}$ are explored by means of the QCD sum rule method. They are
modeled as four-quark mesons $T_{\mathrm{1}}$ and $T_{\mathrm{2}}$ composed
of $b^{T}C\sigma _{\mu \nu }\gamma _{5}b$, $\overline{c}\gamma ^{\nu }C%
\overline{c}^{T}$ and $b^{T}C\gamma _{\mu }\gamma _{5}b$, $\overline{c}C%
\overline{c}^{T}$ diquarks, respectively. The spectroscopic parameters of
the tetraquarks $T_{\mathrm{1}}$ and $T_{\mathrm{2}}$ are determined in the
context of the QCD two-point sum rule method. Results obtained for masses of
these states $m_{1} =(12715\pm 86)~\mathrm{MeV}$ and $m_{2}=(13383\pm 92)~%
\mathrm{MeV}$ are used to fix their strong decay channels. The full width $%
\Gamma (T_{\mathrm{1}})$ of the diquark-antidiquark state $T_{\mathrm{1}}$
is estimated by considering the processes $T_{\mathrm{1}} \to
B_{c}^{-}B_{c}^{\ast -}$ and $T_{\mathrm{1}} \to B_{c}^{\ast -}B_{c}^{\ast
-} $. The decays to mesons $B_{c}^{-}B_{c}^{\ast -}$, $B_{c}^{-}(2S)B_{c}^{%
\ast -}$ and $B_{c}^{\ast -}B_{c}^{\ast -}$ are employed to evaluate $\Gamma
(T_{\mathrm{2}})$. Results obtained for the widths $\Gamma (T_{\mathrm{1}%
})=(44.3\pm 8.8)~\mathrm{MeV}$ and $\Gamma (T_{\mathrm{2}})=(82.5\pm 13.7)~%
\mathrm{MeV}$ of these tetraquarks in conjunction with their masses are
useful for future experimental studies of fully heavy resonances.
\end{abstract}

\maketitle


\section{Introduction}

\label{sec:Intro}
The exotic four-quark mesons containing only heavy $b$ or $c$ quarks and
their properties were already under consideration of researchers. This
interest intensified after discoveries of four $X$ resonances by the LHCb,
ATLAS, and CMS Collaborations \cite%
{LHCb:2020bwg,Bouhova-Thacker:2022vnt,CMS:2023owd}. These new structures
were observed in $J/\psi J/\psi $ and $J/\psi \psi ^{\prime }$ mass
distributions and occupy $6.2-7.3~\mathrm{GeV}$ mass positions in the hadron
spectroscopy. They are presumably fully charmed tetraquarks investigated in
Refs. \cite%
{Zhang:2020xtb,Albuquerque:2020hio,Yang:2020wkh,Becchi:2020mjz,Becchi:2020uvq, Wang:2022xja,Faustov:2022mvs,Niu:2022vqp,Dong:2022sef,Yu:2022lak, Kuang:2023vac,Agaev:2023wua,Agaev:2023ruu,Agaev:2023gaq,Agaev:2023rpj,Agaev:2023ara,Wang:2023kir}
though there are alternative explanations in literature as well \cite%
{Dong:2020nwy,Liang:2021fzr}.

Another class of fully heavy structures are exotic mesons $bb\overline{c}%
\overline{c}$/$cc\overline{b}\overline{b}$. They are interesting objects
because some of them with masses below the relevant $B_{c}B_{c}$ thresholds
may be stable against strong decays. It is clear that such tetraquarks
transform to conventional particles only through electroweak processes, and
should have considerably longer lifetime. In this aspect $bb\overline{c}%
\overline{c}$/$cc\overline{b}\overline{b}$ differ from the tetraquarks $bb%
\overline{b}\overline{b}$ and $cc\overline{c}\overline{c}$:  due
to $b\overline{b}$ or $c\overline{c}$ annihilations they are strong-interaction
unstable even residing below two-particle bottomonia (or charmonia)
production limits \cite{Becchi:2020mjz,Becchi:2020uvq,Agaev:2023ara}.

The next reason to explore the tetraquarks $bb\overline{c}\overline{c}$/$cc%
\overline{b}\overline{b}$ is two units of electric charge that these
particles bear. The existence of double-charged tetraquarks was predicted in
Refs.\ \cite{Chen:2017rhl,Agaev:2017oay}, in which the authors calculated
masses and partial widths of some of their decay modes. Such structures were
analyzed in a detailed form in Refs. \cite{Agaev:2018vag,Agaev:2021jsz} as
well. The first double-charged tetraquark seen in experiment is the scalar
resonance $T_{cs0}^{a}(2900)^{++}$ observed recently by the LHCb
Collaboration \cite{LHCb:2022xob,LHCb:2022bkt}. It has the quark content $cu%
\overline{s}\overline{d}$ and is also a fully open-flavor four-quark system.
It is worth noting that the fully open-flavor tetraquarks were originally
studied in the diquark-antidiquark model in Refs.\ \cite%
{Agaev:2016lkl,Chen:2016mqt}.

The heavy tetraquarks $bb\overline{c}\overline{c}$ were already considered
in the framework of various methods \cite%
{Wu:2016vtq,Li:2019uch,Wang:2019rdo,Liu:2019zuc,Galkin:2023wox,Wang:2021taf,Mutuk:2022nkw}%
, in which the authors investigated different aspects of their physics. In
fact, the masses of particles with $J^{\mathrm{P}}=0^{+}$, $1^{+}$ and $%
2^{+} $ were evaluated in the color-magnetic interaction model in Ref.\ \cite%
{Wu:2016vtq} and found above the $B_{c}B_{c}$ threshold $12549~\mathrm{MeV}$.
Weak decays of the scalar tetraquark $bb\overline{c}\overline{c}$ were
analyzed in Ref.\ \cite{Li:2019uch}. The mass spectra of fully heavy
tetraquark states were calculated in two nonrelativistic quark models which
contain one gluon exchange, linear confinement and hyperfine potentials \cite%
{Wang:2019rdo}. In accordance to these studies the structures $bb\overline{c}%
\overline{c}$ with the quantum numbers $J^{\mathrm{P}}=0^{++}$, $1^{+-}$ and
$2^{++}$ reside above the $B_{c}B_{c}$ thresholds. The masses of the scalar,
axial-vector and tensor tetraquarks $bb\overline{c}\overline{c}$ within a
potential model were found equal to $12947~\mathrm{MeV}$, $12960~\mathrm{MeV}
$ and $12972~\mathrm{MeV}$, respectively, which demonstrate their
strong-interaction unstable nature \cite{Liu:2019zuc}. In the relativistic
quark model,  this problem was addressed recently in Ref.\ \cite{Galkin:2023wox}%
, in which masses of the $J^{\mathrm{P}}=0^{+}$, $1^{+}$ and $2^{+}$
diquark-antidiquark states $bb\overline{c}\overline{c}$ were evaluated as $%
12848~\mathrm{MeV}$, $12852~\mathrm{MeV}$ and $12859~\mathrm{MeV}$,
respectively: All of these particles are above the relevant $B_{c}B_{c}$
thresholds.

But analyses carried out using QCD moment sum rule method led to conclusions
that certain structures $cc\overline{b}\overline{b}$ with $J^{\mathrm{P}%
}=0^{+}$, $1^{+}$ and $2^{+}$ are strong-interaction stable states \ \cite%
{Wang:2021taf}. As a result, such tetraquarks can transform to ordinary
particles only through electroweak processes. The same structures were
investigated in the context of the dynamical diquark model and predicted to
be stable against strong decays as well \cite{Mutuk:2022nkw}.

In our article \cite{Agaev:2023tzi}, we explored the scalar tetraquarks $X_{%
\mathrm{1}}$,$X_{\mathrm{2}}=bb\overline{c}\overline{c}$ composed of
axial-vector and pseudoscalar diquarks, respectively. The masses of these
states $(12715\pm 80)~\mathrm{MeV}$ and $(13370\pm 95)~\mathrm{MeV}$
overshoot the threshold for production of $B_{c}^{-}B_{c}^{-}$ mesons. This
means that the scalar tetraquarks $X_{\mathrm{1}}$ and $X_{\mathrm{2}}$ with
different internal organizations are strong-interaction unstable particles.
We evaluated also their full widths by considering the strong decay channels
$X_{\mathrm{1}}\rightarrow B_{c}^{-}B_{c}^{-}$ and\ $B_{c}^{\ast
-}B_{c}^{\ast -}$, and $X_{\mathrm{2}}\rightarrow B_{c}^{-}B_{c}^{-}$,\ $%
B_{c}^{-}B_{c}^{-}(2S)$, and$\ B_{c}^{\ast -}B_{c}^{\ast -}$, respectively.
Our predictions for the full widths of these states $\Gamma _{\mathrm{1}%
}=(63\pm 12)~\mathrm{MeV}$ and $\Gamma _{\mathrm{2}}=(79\pm 14)~\mathrm{MeV}$
allowed us to interpret them as tetraquarks with moderate width.

In the present article, we continue our investigations of the fully heavy
tetraquarks by analyzing the axial-vector particles $bb\overline{c}\overline{%
c}$. We model the tetraquark $T_{\mathrm{1}}$ as an exotic meson with
structure $C\sigma _{\mu \nu }\gamma _{5}\otimes \gamma ^{\nu }C$ built of a
diquark $b^{T}C\sigma _{\mu \nu }\gamma _{5}b$ and antidiquark $\overline{c}%
\gamma ^{\nu }C\overline{c}^{T}$. The tetraquark $T_{\mathrm{2}}$ has the
composition $C\gamma _{\mu }\gamma _{5}\otimes C$ and contains a vector
diquark and pseudoscalar antidiquark. The states $T_{\mathrm{1}}$ and $T_{%
\mathrm{2}}$ have color triplet $\overline{\mathbf{3}_{c}}\otimes \mathbf{3}%
_{c}{}$ and sextet $\mathbf{6}_{c}\otimes \overline{\mathbf{6}}_{c}$
structures, respectively. The spectroscopic parameters of these tetraquarks
are computed in the framework of the QCD sum rule (SR) approach \cite%
{Shifman:1978bx,Shifman:1978by}. To find widths of $T_{\mathrm{1}}$ and $T_{%
\mathrm{2}}$, we invoke the three-point SR method which is required to
evaluate the strong couplings of $T_{\mathrm{1(2)}}$ and final state-mesons $%
B_{c}B_{c}$ with appropriate spin-parities and charges.

This work is composed in the following manner: In Sec.\ \ref{sec:AV}, we
compute the spectroscopic parameters of the axial-vector tetraquarks $T_{%
\mathrm{1}}$ and $T_{\mathrm{2}}$. Using obtained predictions for masses of
these systems, we determine their kinematically allowed decay channels. In
Sec.\ \ref{sec:AVWidths1}, we consider decay modes of $T_{\mathrm{1}}$ and
compute its full width. The section\ \ref{sec:AVWidths2} is devoted to
analysis of decays of the structure $T_{\mathrm{2}}$. Our concluding notes
are presented in Sec.\ \ref{sec:Conc}.


\section{Mass and current coupling of $T_{\mathrm{1}}$ and $T_{\mathrm{2}}$}

\label{sec:AV} 

In this section, we calculate the mass $m_{1(2)}$ and current coupling $%
\Lambda _{1(2)}$ (a pole residue) of the axial-vector tetraquark $T_{\mathrm{%
1(2)}}=bb\overline{c}\overline{c}$ in the QCD two-point SR framework.

The SRs for the quantities $m_{1}$, $\Lambda _{1}$ and $m_{2}$, $\Lambda
_{2} $ can be derived from analysis of the following correlator%
\begin{equation}
\Pi _{\mu \nu }(p)=i\int d^{4}xe^{ipx}\langle 0|\mathcal{T}\{J_{\mu
}(x)J_{\nu }^{\dag }(0)\}|0\rangle .  \label{eq:CF1}
\end{equation}%
Here $J_{\mu }(x)$ is an interpolating current of the particle under
consideration. The symbol $\mathcal{T}$ is adopted for the time-ordering of
two currents. To model the tetraquarks $T_{\mathrm{1}}$ and $T_{\mathrm{2}}$%
, we use the currents
\begin{equation}
J_{\mu }^{1}(x)=[b_{a}^{T}(x)C\sigma _{\mu \nu }\gamma _{5}b_{b}(x)][%
\overline{c}_{a}(x)\gamma ^{\nu }C\overline{c}_{b}^{T}(x)],  \label{eq:CR1}
\end{equation}%
and
\begin{equation}
J_{\mu }^{2}(x)=[b_{a}^{T}(x)C\gamma _{\mu }\gamma _{5}b_{b}(x)][\overline{c}%
_{a}(x)C\overline{c}_{b}^{T}(x)].  \label{eq:CR2}
\end{equation}

Let us consider in a detailed form computation of the parameters $m_{1}$ and
$\Lambda _{1}$. The correlation function $\Pi _{\mu \nu }^{1}(p)$ can be
presented using the physical parameters of the tetraquark $T_{\mathrm{1}}$.
Having inserted into Eq.\ (\ref{eq:CF1}) a full set of states with the
spin-parities and contents of $T_{\mathrm{1}}$, and carried out integration
over $x$, we get
\begin{equation}
\Pi _{\mu \nu }^{1\mathrm{Phys}}(p)=\frac{\langle 0|J_{\mu }^{1}|T_{\mathrm{1%
}}(p,\epsilon )\rangle \langle T_{\mathrm{1}}(p,\epsilon
)|J_{v}^{1}{}^{\dagger }|0\rangle }{m_{1}^{2}-p^{2}}+\cdots .
\label{eq:Phys1}
\end{equation}%
In expression above, only the contribution of the ground-state particle is
written down explicitly: Effects of higher resonances and continuum states
are denoted by the dots.

For further detailing of $\Pi _{\mu \nu }^{1\mathrm{Phys}}(p)$, it is
convenient to introduce the matrix element
\begin{equation}
\langle 0|J_{\mu }^{1}|T_{\mathrm{1}}(p,\epsilon )\rangle =\Lambda
_{1}\epsilon _{\mu }(p).  \label{eq:ME1}
\end{equation}%
where $\epsilon _{\mu }$ is the polarization vector of $T_{\mathrm{1}}$.
Then, it is easy to find $\Pi _{\mu \nu }^{1\mathrm{Phys}}(p)$ which is
given by the formula
\begin{equation}
\Pi _{\mu \nu }^{1\mathrm{Phys}}(p)=\frac{\Lambda _{1}^{2}}{m_{1}^{2}-p^{2}}%
\left( -g_{\mu \nu }+\frac{p_{\mu }p_{\nu }}{m^{2}}\right) +\cdots .
\label{eq:PhysSide1}
\end{equation}%
In what follows, we are going to use the invariant amplitude $\Pi ^{1\mathrm{%
Phys}}(p^{2})$ corresponding to the structure $g_{\mu \nu }$.

The QCD side of the SRs for the parameters $m_{1}$ and $\Lambda _{1}$ is
equal to
\begin{eqnarray}
&&\Pi _{\mu \nu }^{1\mathrm{OPE}}(p)=i\int d^{4}xe^{ipx}\left\{ \mathrm{Tr}%
\left[ \gamma ^{\theta }\widetilde{S}_{c}^{b^{\prime }b}(-x)\gamma ^{\delta
}S_{c}^{a^{\prime }a}(-x)\right] \right.  \notag \\
&&\times \left[ \mathrm{Tr}\left[ S_{b}^{bb^{\prime }}(x)\gamma _{5}\sigma
_{\nu \delta }\widetilde{S}_{b}^{aa^{\prime }}(x)\sigma _{\mu \theta }\gamma
_{5}\right] \right.  \notag \\
&&\left. -\mathrm{Tr}\left[ S_{b}^{ab^{\prime }}(x)\gamma _{5}\sigma _{\nu
\delta }\widetilde{S}_{b}^{ba^{\prime }}(x)\sigma _{\mu \theta }\gamma _{5}%
\right] \right] +\mathrm{Tr}\left[ \gamma ^{\theta }\widetilde{S}%
_{c}^{a^{\prime }b}(-x)\right.  \notag \\
&&\left. \times \gamma ^{\delta }S_{c}^{b^{\prime }a}(-x)\right] \left[
\mathrm{Tr}\left[ S_{b}^{ab^{\prime }}(x)\gamma _{5}\sigma _{\nu \delta }%
\widetilde{S}_{b}^{ba^{\prime }}(x)\sigma _{\mu \theta }\gamma _{5}\right]
\right.  \notag \\
&&\left. \left. -\mathrm{Tr}\left[ S_{b}^{bb^{\prime }}(x)\gamma _{5}\sigma
_{\nu \delta }\widetilde{S}_{b}^{aa^{\prime }}(x)\sigma _{\mu \theta }\gamma
_{5}\right] \right] \right\} ,  \label{eq:QCD1}
\end{eqnarray}%
where $\widetilde{S}_{Q}(x)=CS_{Q}^{T}(x)C$ with $S_{b(c)}(x)$ being the $b$
and $c$-quark propagators \cite{Agaev:2020zad}. We denote by $\Pi ^{1\mathrm{%
OPE}}(p^{2})$ the invariant amplitude which corresponds to the term $g_{\mu
\nu }$ in the correlator $\Pi _{\mu \nu }^{1\mathrm{OPE}}(p)$.

The sum rules for the mass and current coupling of the tetraquark $T_{%
\mathrm{1}}$ are given by the expressions%
\begin{equation}
m_{1}^{2}=\frac{\Pi ^{1\prime }(M^{2},s_{0})}{\Pi ^{1}(M^{2},s_{0})}
\label{eq:Mass}
\end{equation}%
and
\begin{equation}
\Lambda _{1}^{2}=e^{m_{1}^{2}/M^{2}}\Pi ^{1}(M^{2},s_{0}),  \label{eq:Coupl}
\end{equation}%
where $\Pi ^{1}(M^{2},s_{0})$ is the amplitude $\Pi ^{1\mathrm{OPE}}(p^{2})$
after the Borel transformation and continuum subtraction procedures. Here $%
M^{2}$ and $s_{0}$ are the Borel and continuum subtraction parameters,
respectively. In Eq.\ (\ref{eq:Mass}), we also adopt a notation $\Pi
^{1\prime }(M^{2},s_{0})=d\Pi ^{1}(M^{2},s_{0})/d(-1/M^{2})$.

\begin{figure}[h]
\includegraphics[width=8.5cm]{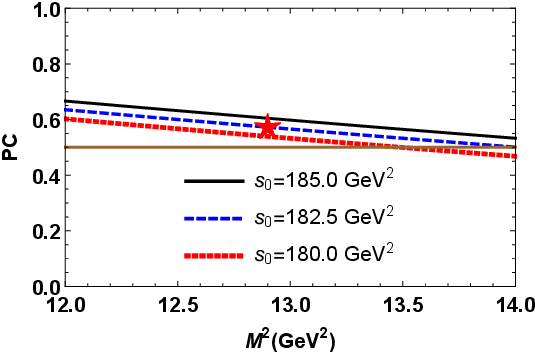}
\caption{The pole contribution $\mathrm{PC}$ as a function of the Borel
parameter $M^{2}$ at different $s_{0}$. The horizontal line limits a region $%
\mathrm{PC}=0.5$. The red star marks the point, where the mass $m $ of $T_{%
\mathrm{1}}$ has effectively been computed. }
\label{fig:PC}
\end{figure}

In numerical computations, we employ the input parameters
\begin{eqnarray}
\langle \alpha _{s}G^{2}/\pi \rangle &=&(0.012\pm 0.004)~\mathrm{GeV}^{4},
\notag \\
m_{b} &=&4.18_{-0.02}^{+0.03}~\mathrm{GeV},  \notag \\
\ m_{c} &=&(1.27\pm 0.02)~\mathrm{GeV.}  \label{eq:Param}
\end{eqnarray}%
The auxiliary quantities $M^{2}$ and $s_{0}$ are chosen within limits
\begin{equation}
M^{2}\in \lbrack 12,14]~\mathrm{GeV}^{2},\ s_{0}\in \lbrack 180,185]~\mathrm{%
GeV}^{2},  \label{eq:Wind1}
\end{equation}%
which comply with all constraints of SR calculations. Indeed, at $M^{2}=14~%
\mathrm{GeV}^{2}$ and $M^{2}=12~\mathrm{GeV}^{2}$ on the average in $s_{0}$
the pole contribution is $\mathrm{PC}\approx 0.5$ and $\mathrm{PC}$ $\approx
0.64$, respectively (see, Fig.\ \ref{fig:PC}). At $M^{2}=12~\mathrm{GeV}^{2}$
the nonperturbative contribution is positive and forms $2\%$ of the whole
result.

The mass $m_{1}$ and current coupling $\Lambda _{1}$ are found as mean
values of these parameters calculated at $10$ different points from the
regions Eq.\ (\ref{eq:Wind1}). For example, the maximum $m_{1}=12800~\mathrm{%
MeV}$ is reached at $M^{2}=14\ \mathrm{GeV}^{2}$ and $s_{0}=185~\mathrm{GeV}%
^{2}$, whereas $m_{1}$ gets its minimum value $12630~\mathrm{MeV}$ at $%
M^{2}=12\ \mathrm{GeV}^{2}$ and $s_{0}=180~\mathrm{GeV}^{2}$. As a result,
we find
\begin{eqnarray}
m_{1} &=&(12714\pm 86)~\mathrm{MeV},  \notag \\
\Lambda _{1} &=&(2.27\pm 0.26)~\mathrm{GeV}^{5}.  \label{eq:Result1}
\end{eqnarray}%
Here, the errors are generated mainly by uncertainties in the choice of the
parameters $M^{2}$ and $s_{0}$: Effects of ambiguities in Eq.\ (\ref%
{eq:Param}) on the final output are  small. The results in Eq.\ (\ref%
{eq:Result1}) effectively amount to the SR predictions at the point $%
M^{2}=12.9~\mathrm{GeV}^{2}$ and $s_{0}=182.5~\mathrm{GeV}^{2}$, where the
pole contribution is $\mathrm{PC}\approx 0.57$. This fact guarantees the
dominance of $\mathrm{PC}$ in the obtained results, and proves ground-level
nature of $T_{\mathrm{1}}$ in segment of axial-vector tetraquarks $bb%
\overline{c}\overline{c}$. The dependence of the mass $m_{1}$ on $M^{2}$ and
$s_{0}$ is shown in Fig.\ \ref{fig:Mass1}.

\begin{widetext}

\begin{figure}[h!]
\begin{center}
\includegraphics[totalheight=6cm,width=8cm]{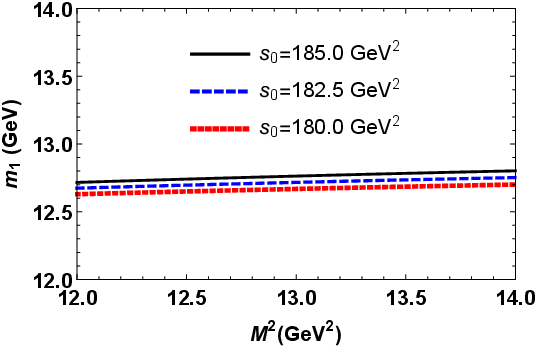}
\includegraphics[totalheight=6cm,width=8cm]{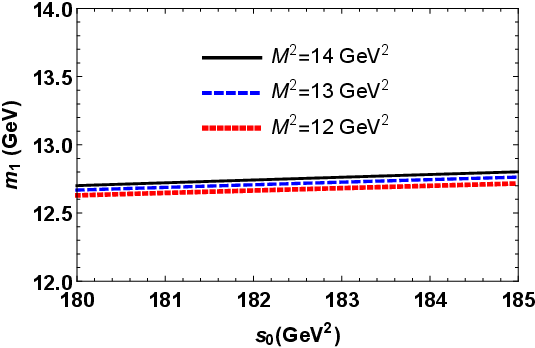}
\end{center}
\caption{Mass $m_1$ of the tetraquark $T_{\mathrm 1}$ as a function of the Borel  $M^{2}$ (left panel), and continuum threshold $s_0$ parameters (right panel).}
\label{fig:Mass1}
\end{figure}

\begin{figure}[h!]
\begin{center}
\includegraphics[totalheight=6cm,width=8cm]{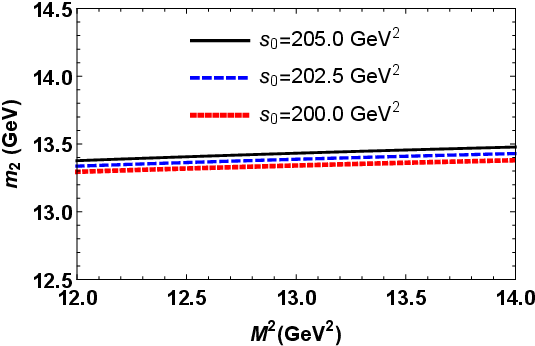}
\includegraphics[totalheight=6cm,width=8cm]{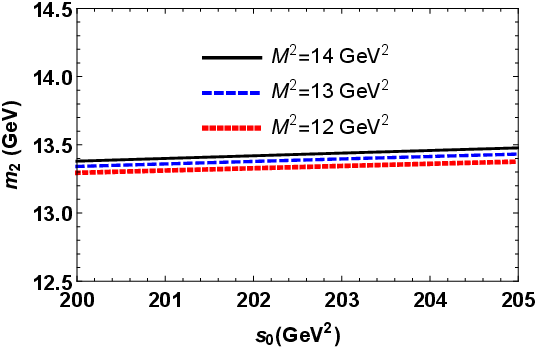}
\end{center}
\caption{Dependence of the mass $m_2$ on the Borel parameter $M^{2}$ (left panel), and continuum threshold parameter $s_0$ (right panel).}
\label{fig:Mass2}
\end{figure}

\end{widetext}

The parameters of the tetraquark $T_{\mathrm{2}}$ are determined by the
similar manner. Below, we provide the correlation function $\Pi _{\mu \nu
}^{2\mathrm{OPE}}(p)$ for the current $J_{\mu }^{2}(x)$
\begin{eqnarray}
&&\Pi _{\mu \nu }^{2\mathrm{OPE}}(p)=i\int d^{4}xe^{ipx}\left\{ \mathrm{Tr}%
\left[ \widetilde{S}_{c}^{b^{\prime }b}(-x)S_{c}^{a^{\prime }a}(-x)\right]
\right.  \notag \\
&&\times \left[ \mathrm{Tr}\left[ S_{b}^{bb^{\prime }}(x)\gamma _{5}\gamma
_{\nu }\widetilde{S}_{b}^{aa^{\prime }}(x)\gamma _{\mu }\gamma _{5}\right]
\right.  \notag \\
&&\left. +\mathrm{Tr}\left[ S_{b}^{ab^{\prime }}(x)\gamma _{5}\gamma _{\nu }%
\widetilde{S}_{b}^{ba^{\prime }}(x)\gamma _{\mu }\gamma _{5}\right] \right] +%
\mathrm{Tr}\left[ S_{c}^{b^{\prime }a}(-x)\right.  \notag \\
&&\left. \times \widetilde{S}_{c}^{a^{\prime }b}(-x)\right] \left[ \mathrm{Tr%
}\left[ S_{b}^{bb^{\prime }}(x)\gamma _{5}\gamma _{\nu }\widetilde{S}%
_{b}^{aa^{\prime }}(x)\gamma _{\mu }\gamma _{5}\right] \right.  \notag \\
&&\left. \left. +\mathrm{Tr}\left[ S_{b}^{ab^{\prime }}(x)\gamma _{5}\gamma
_{\nu }\widetilde{S}_{b}^{ba^{\prime }}(x)\gamma _{\mu }\gamma _{5}\right] %
\right] \right\} .  \label{eq:QCD2}
\end{eqnarray}%
The spectroscopic parameters $m_{2}$ and $\Lambda _{2}$ of this particle are
equal to
\begin{eqnarray}
m_{2} &=&(13383\pm 92)~\mathrm{MeV},  \notag \\
\Lambda _{2} &=&(1.01\pm 0.12)~\mathrm{GeV}^{5},  \label{eq:Result2}
\end{eqnarray}
which are obtained by employing the working windows
\begin{equation}
M^{2}\in \lbrack 12,14]~\mathrm{GeV}^{2},\ s_{0}\in \lbrack 200,205]~\mathrm{%
GeV}^{2}.  \label{eq:Wind2}
\end{equation}%
The pole contribution for $M^{2}\in \lbrack 12,14]~\mathrm{GeV}^{2}$ on
average in $s_{0}$ is larger than $0.5$ and varies within limits
\begin{equation}
0.65\geq \mathrm{PC}\geq 0.50.
\end{equation}%
At $M^{2}=12~\mathrm{GeV}^{2}$ the nonperturbative contribution is negative
and does not exceed $3\%$ of the correlation function. Our predictions for
the mass $m_{2}$ of the tetraquark $T_{\mathrm{2}}$ are plotted in Fig.\ \ref%
{fig:Mass2}.


\section{Width of the tetraquark $T_{\mathrm{1}}$}

\label{sec:AVWidths1}


The mass $m_{1}$ of the diquark-antidiquark state $T_{\mathrm{1}}$ proves
that it can decay to $B_{c}^{-}B_{c}^{\ast -}$ and $B_{c}^{\ast
-}B_{c}^{\ast -}$mesons. In fact, the mass $m_{B_{c}}=(6274.47\pm 0.27)~%
\mathrm{MeV}$ of the meson $B_{c}^{-}$ is known experimentally \cite%
{PDG:2022}. For the mass of the vector particle $B_{c}^{\ast -}$, we use the
theoretical prediction $m_{B_{c}^{\ast }}=6338~\mathrm{MeV}$ from Ref.\ \cite%
{Godfrey:2004ya}. It is evident, that processes $T_{\mathrm{1}}\rightarrow
B_{c}^{-}B_{c}^{\ast -}$ and $T_{\mathrm{1}}\rightarrow B_{c}^{\ast
-}B_{c}^{\ast -}$ are kinematically allowed channels for the tetraquark $T_{%
\mathrm{1}}$. As the decay constants of $B_{c}^{-}$ and $B_{c}^{\ast -}$
mesons, we employ $f_{B_{c}}=(476\pm 27)~\mathrm{MeV}$ and $f_{B_{c}^{\ast
}}=471~\mathrm{MeV}$ \cite{Veliev:2010vd,Eichten:2019gig}, respectively.

\subsection{Process $T_{\mathrm{1}}\rightarrow B_{c}^{-}B_{c}^{\ast -}$}


The width of the decay $T_{\mathrm{1}}\rightarrow B_{c}^{-}B_{c}^{\ast -}$
can be calculated by employing the strong coupling $f_{1}$ of particles at
the vertex $T_{\mathrm{1}}B_{c}^{-}B_{c}^{\ast -}$. To this end, we are
going to study the QCD three-point correlation function
\begin{eqnarray}
\Pi _{\mu \nu }^{1}(p,p^{\prime }) &=&i^{2}\int d^{4}xd^{4}ye^{ip^{\prime
}y}e^{-ipx}\langle 0|\mathcal{T}\{J_{\mu }^{B_{c}^{\ast }}(y)  \notag \\
&&\times J^{B_{c}}(0)J_{\nu }^{1\dag }(x)\}|0\rangle ,  \label{eq:CF3}
\end{eqnarray}%
where
\begin{eqnarray}
J^{B_{c}}(x) &=&\overline{c}_{j}(x)i\gamma _{5}b_{j}(x),  \notag \\
J_{\mu }^{B_{c}^{\ast }}(x) &=&\overline{c}_{i}(x)\gamma _{\mu }b_{i}(x)
\label{eq:CR3}
\end{eqnarray}%
are the interpolating currents for the pseudoscalar and vector mesons $%
B_{c}^{-}$ and $B_{c}^{\ast -}$, respectively.

Our aim is to analyze the correlator $\Pi _{\mu \nu }^{1}(p,p^{\prime })$
and find the sum rule for the form factor $f_{1}(q^{2})$ which at $%
q^{2}=m_{B_{c}}^{2}$ gives the strong coupling $f_{1}$. The SR for $%
f_{1}(q^{2})$ can be derived by means of usual recipes of the approach.
First, we represent $\Pi _{\mu \nu }^{1}(p,p^{\prime })$ by employing the
parameters of $T_{\mathrm{1}}$ and final-state mesons. The correlation
function $\Pi _{\mu \nu }^{1\mathrm{Phys}}(p,p^{\prime })$ calculated by
this way establishes the phenomenological side of SR and has the form
\begin{eqnarray}
&&\Pi _{\mu \nu }^{1\mathrm{Phys}}(p,p^{\prime })=\frac{\langle 0|J_{\mu
}^{B_{c}^{\ast }}|B_{c}^{\ast }(p^{\prime },\varepsilon )\rangle }{p^{\prime
2}-m_{B_{c}^{\ast }}^{2}}\frac{\langle 0|J^{B_{c}}|B_{c}(q)\rangle }{%
q^{2}-m_{B_{c}}^{2}}  \notag \\
&&\times \langle B_{c}^{\ast }(p^{\prime },\varepsilon )B_{c}(q)|T_{\mathrm{1%
}}(p,\epsilon )\rangle \frac{\langle T_{\mathrm{1}}(p,\epsilon )|J_{\nu
}^{1\dag }|0\rangle }{p^{2}-m_{1}^{2}}  \notag \\
&&+\cdots .  \label{eq:CF5}
\end{eqnarray}

To rewrite $\Pi _{\mu \nu }^{1\mathrm{Phys}}(p,p^{\prime })$ in a form
suitable for further manipulations, we make use of the matrix elements
\begin{eqnarray}
\langle 0|J^{B_{c}}|B_{c}\rangle &=&\frac{f_{B_{c}}m_{B_{c}}^{2}}{m_{b}+m_{c}%
},  \notag \\
\langle 0|J_{\mu }^{B_{c}^{\ast }}|B_{c}^{\ast }(p^{\prime },\varepsilon
)\rangle &=&f_{B_{c}^{\ast }}m_{B_{c}^{\ast }}\varepsilon _{\mu }(p^{\prime
}),  \label{eq:ME2}
\end{eqnarray}%
where $\varepsilon _{\mu }$ is the polarization vector of the meson $%
B_{c}^{\ast -}$. It is convenient to model the vertex $T_{\mathrm{1}%
}B_{c}^{-}B_{c}^{\ast -}$ in the following form
\begin{eqnarray}
&&\langle B_{c}^{\ast }(p^{\prime },\varepsilon )B_{c}(q)|T_{\mathrm{1}%
}(p,\epsilon )\rangle =f_{1}(q^{2})\left[ (p\cdot p^{\prime })(\epsilon
\cdot \varepsilon ^{\ast })\right.  \notag \\
&&\left. -(p^{\prime }\cdot \epsilon )(p\cdot \varepsilon ^{\ast }\right] .
\label{eq:ME3}
\end{eqnarray}%
Having determined the required matrix elements, it is not difficult to find
that
\begin{eqnarray}
&&\Pi _{\mu \nu }^{1\mathrm{Phys}}(p,p^{\prime })=f_{1}(q^{2})\frac{\Lambda
_{1}f_{B_{c}}m_{B_{c}}^{2}f_{B_{c}^{\ast }}m_{B_{c}^{\ast }}}{%
(m_{b}+m_{c})\left( p^{2}-m_{1}^{2}\right) \left( p^{\prime
2}-m_{B_{c}^{\ast }}^{2}\right) }  \notag \\
&&\times \frac{1}{(q^{2}-m_{B_{c}}^{2})}\left( \frac{m_{1}^{2}+m_{B_{c}^{%
\ast }}^{2}-q^{2}}{2}g_{\mu \nu }-p_{\mu }p_{\nu }^{\prime }\right) +\cdots .
\label{eq:CF6}
\end{eqnarray}%
The correlator $\Pi _{\mu \nu }^{1\mathrm{Phys}}(p,p^{\prime })$ is a sum of
two different Lorentz structures, one of which should be chosen for
investigations. We work with the invariant amplitude $\Pi _{1}^{\mathrm{Phys}%
}(p^{2},p^{\prime 2},q^{2})$ that corresponds, in Eq.\ (\ref{eq:CF6}), to
the structure $g_{\mu \nu }$.

The correlator $\Pi _{\mu \nu }^{1\mathrm{OPE}}(p,p^{\prime })$, expressed
using the quark propagators, reads
\begin{eqnarray}
&&\Pi _{\mu \nu }^{1\mathrm{OPE}}(p,p^{\prime })=2i^{3}\int
d^{4}xd^{4}ye^{ip^{\prime }y}e^{-ipx}\left\{ \mathrm{Tr}\left[ \gamma
_{5}S_{b}^{ia}(-x)\right. \right.  \notag \\
&&\left. \times \gamma _{5}\sigma _{\nu \theta }\widetilde{S}%
_{b}^{jb}(y-x)\gamma _{\mu }\widetilde{S}_{c}^{aj}(x-y)\gamma ^{\theta
}S_{c}^{bi}(x)\right]  \notag \\
&&-\mathrm{Tr}\left[ \gamma _{5}S_{b}^{ia}(-x)\gamma _{5}\sigma _{\nu \theta
}\widetilde{S}_{b}^{jb}(y-x)\gamma _{\mu }\widetilde{S}_{c}^{bj}(x-y)\right.
\notag \\
&&\left. \left. \times \gamma ^{\theta }S_{c}^{ai}(x)\right] \right\} .
\label{eq:QCDside2}
\end{eqnarray}%
The $\Pi _{\mu \nu }^{1\mathrm{OPE}}(p,p^{\prime })$ also consists of two
Lorentz structures proportional to $g_{\mu \nu }$ and $p_{\mu }p_{\nu
}^{\prime }$, respectively. Having labeled by $\Pi _{1}^{\mathrm{OPE}%
}(p^{2},p^{\prime 2},q^{2})$ the amplitude corresponding to the term $g_{\mu
\nu }$, we determine the SR for the form factor $f_{1}(q^{2})$
\begin{eqnarray}
&&f_{1}(q^{2})=\frac{2(m_{b}+m_{c})}{\Lambda
_{1}f_{B_{c}}m_{B_{c}}^{2}f_{B_{c}^{\ast }}m_{B_{c}^{\ast }}}\frac{%
q^{2}-m_{B_{c}}^{2}}{m^{2}+m_{B_{c}^{\ast }}^{2}-q^{2}}  \notag \\
&&\times e^{m_{1}^{2}/M_{1}^{2}}e^{m_{B_{c}^{\ast }}^{2}/M_{2}^{2}}\Pi _{1}(%
\mathbf{M}^{2},\mathbf{s}_{0},q^{2}).  \label{eq:SRCoup2}
\end{eqnarray}%
In Eq.\ (\ref{eq:SRCoup2}), $\Pi _{1}(\mathbf{M}^{2},\mathbf{s}_{0},q^{2})$
is the function $\Pi _{1}^{\mathrm{OPE}}(p^{2},p^{\prime 2},q^{2})$ after
the Borel transformations and continuum subtractions. As a result, it
depends on the parameters $\mathbf{M}^{2}=(M_{1}^{2},M_{2}^{2})$ and $%
\mathbf{s}_{0}=(s_{0},s_{0}^{\prime })$. The pair $(M_{1}^{2},s_{0})$
corresponds to the initial tetraquark channel, whereas $(M_{2}^{2},s_{0}^{%
\prime })$ describes the $B_{c}^{\ast -}$ channel.

In numerical computations for $M_{1}^{2}$ and $s_{0}$, we use Eq.\ (\ref%
{eq:Wind1}). The parameters $(M_{2}^{2},\ s_{0}^{\prime })$ for the $%
B_{c}^{\ast -}$ channel are varied inside of the borders%
\begin{equation}
M_{2}^{2}\in \lbrack 6.5,7.5]~\mathrm{GeV}^{2},\ s_{0}^{\prime }\in \lbrack
49,51]~\mathrm{GeV}^{2}.  \label{eq:Wind3}
\end{equation}

The sum rule method leads to credible results for the form factor $%
f_{1}(q^{2})$ in the Euclidean region $q^{2}<0$. But the strong coupling $%
f_{1}$ is determined by $f_{1}(q^{2})$ at the mass shell $%
q^{2}=m_{B_{c}}^{2} $. To solve this problem, it is convenient to use the
function $f_{1}(Q^{2})$ with $Q^{2}=-q^{2}$ and introduce a fit function $%
\mathcal{F}_{1}(Q^{2},m_{1}^{2})=f_{1}(Q^{2})$ that at momenta $Q^{2}>0$
coincides with SR data, but can be extended to the domain $Q^{2}<0$. To this
end, we employ the functions
\begin{equation}
\mathcal{F}_{i}(Q^{2},m_{1}^{2})=\mathcal{F}_{i}^{0}\mathrm{\exp }\left[
c_{i}^{1}\frac{Q^{2}}{m_{1}^{2}}+c_{i}^{2}\left( \frac{Q^{2}}{m_{1}^{2}}%
\right) ^{2}\right]  \label{eq:FitF}
\end{equation}%
where $\mathcal{F}_{i}^{0}$, $c_{i}^{1}$, and $c_{i}^{2}$ are unknown
parameters.

\begin{figure}[h]
\includegraphics[width=8.5cm]{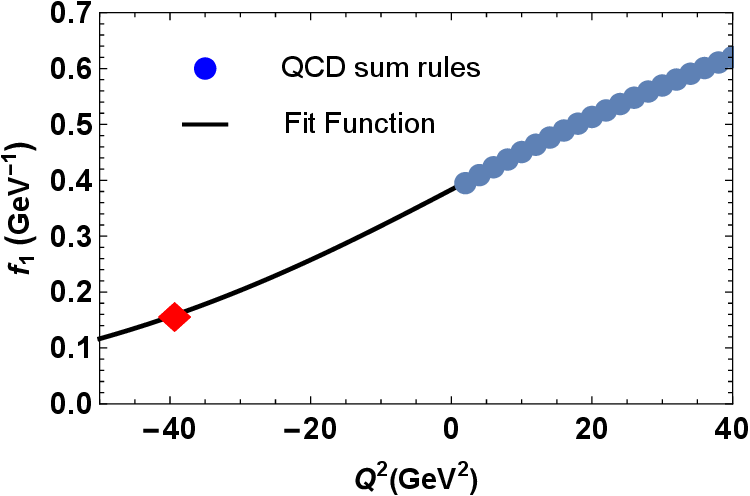}
\caption{QCD data and fit function for the form factor $f_{1}(Q^{2})$. The
diamond fixes the point $Q^{2}=-m_{B_{c}}^{2}$ where the coupling $f_{1}$
has been evaluated. }
\label{fig:Fit}
\end{figure}

In the present SR computations, $Q^{2}$ changes within the interval $%
Q^{2}=1-40~\mathrm{GeV}^{2}$. The results obtained for $f_{1}(Q^{2})$ are
depicted in Fig.\ \ref{fig:Fit}. Then, by comparing QCD data and Eq.\ (\ref%
{eq:FitF}), it is not difficult to extract the parameters $\mathcal{F}%
_{1}^{0}=0.38~\mathrm{GeV}^{-1}$, $c_{1}^{1}=2.79$, and $c_{1}^{2}=-3.44$ of
the function $\mathcal{F}_{1}(Q^{2},m_{1}^{2})$. It is also plotted in Fig.\ %
\ref{fig:Fit}, where one sees a nice agreement of $\mathcal{F}%
_{1}(Q^{2},m_{1}^{2})$ with QCD data.

For the strong coupling $f_{1}$, we find
\begin{equation}
f_{1}\equiv \mathcal{F}_{1}(-m_{B_{c}}^{2},m_{1}^{2})=(1.6\pm 0.2)\times
10^{-1}\ \mathrm{GeV}^{-1}.
\end{equation}%
The width of the process $T_{\mathrm{1}}\rightarrow B_{c}^{-}B_{c}^{\ast -}$
is determined by the expression%
\begin{equation}
\Gamma \left[ T_{\mathrm{1}}\rightarrow B_{c}^{-}B_{c}^{\ast -}\right]
=f_{1}^{2}\frac{m_{B_{c}^{\ast }}^{2}\lambda _{1}}{24\pi }\left( 3+\frac{%
2\lambda _{1}^{2}}{m_{B_{c}^{\ast }}^{2}}\right) ,  \label{eq:PDw2}
\end{equation}%
where $\lambda _{1}=\lambda (m_{1},m_{B_{c}^{\ast }},m_{B_{c}})$, and
\begin{equation}
\lambda (x,y,z)=\frac{\sqrt{%
x^{4}+y^{4}+z^{4}-2(x^{2}y^{2}+x^{2}z^{2}+y^{2}z^{2})}}{2x}.
\end{equation}%
As a result, we find
\begin{equation}
\Gamma \left[ T_{\mathrm{1}}\rightarrow B_{c}^{-}B_{c}^{\ast -}\right]
=(32.5\pm 8.3)~\mathrm{MeV}.  \label{eq:DW1}
\end{equation}

\subsection{Decay $T_{\mathrm{1}}\rightarrow B_{c}^{\ast -}B_{c}^{\ast -}$}


To explore the process $T_{\mathrm{1}}\rightarrow B_{c}^{\ast -}B_{c}^{\ast
-}$, we start from the correlation function
\begin{eqnarray}
\Pi _{\mu \delta \nu }^{1}(p,p^{\prime }) &=&i^{2}\int
d^{4}xd^{4}ye^{ip^{\prime }y}e^{-ipx}\langle 0|\mathcal{T}\{J_{\mu
}^{B_{c}^{\ast }}(y)  \notag \\
&&\times J_{\delta }^{B_{c}^{\ast }}(0)J_{\nu }^{1\dag }(x)\}|0\rangle .
\label{eq:CF7}
\end{eqnarray}%
To derive expression of this function in terms of physical parameters of the
tetraquark $T_{\mathrm{1}}$ and $B_{c}^{\ast -}$ meson, we write it in the
following form
\begin{eqnarray}
&&\Pi _{\mu \delta \nu }^{1\mathrm{Phys}}(p,p^{\prime })=\frac{\langle
0|J_{\mu }^{B_{c}^{\ast }}|B_{c}^{\ast }(p^{\prime },\varepsilon (p^{\prime
}))\rangle }{p^{\prime 2}-m_{B_{c}^{\ast }}^{2}}\frac{\langle 0|J_{\delta
}^{B_{c}^{\ast }}|B_{c}^{\ast }(q,\varepsilon (q))\rangle }{%
q^{2}-m_{B_{c}^{\ast }}^{2}}  \notag \\
&&\times \langle B_{c}^{\ast }(p^{\prime },\varepsilon (p^{\prime
}))B_{c}^{\ast }(q,\varepsilon (q))|T_{\mathrm{1}}(p,\epsilon )\rangle \frac{%
\langle T_{\mathrm{1}}(p,\epsilon )|J_{\nu }^{1\dag }|0\rangle }{%
p^{2}-m_{1}^{2}}  \notag \\
&&+\cdots .  \label{eq:PhysSide2}
\end{eqnarray}%
The matrix elements of the $B_{c}^{\ast -}$ meson and diquark-antidiquark
state $T_{\mathrm{1}}$ have been defined above. An unknown matrix element
here is one connected with the vertex $T_{\mathrm{1}}B_{c}^{\ast
-}B_{c}^{\ast -}$, which we model by means of the expression%
\begin{eqnarray}
&&\langle B_{c}^{\ast }(p^{\prime },\varepsilon (p^{\prime }))B_{c}^{\ast
}(q,\varepsilon (q))|T_{\mathrm{1}}(p,\epsilon )\rangle
=f_{2}(q^{2})\epsilon ^{\alpha \beta \gamma \zeta }  \notag \\
&&\times \varepsilon _{\alpha }^{\ast }(p^{\prime })\epsilon _{\beta
}(p)\varepsilon _{\gamma }^{\ast }(q)(p_{\zeta }^{\prime }+p_{\zeta }).
\label{eq:Mel}
\end{eqnarray}

The correlator $\Pi _{\mu \gamma \nu }^{1}(p,p^{\prime })$ in terms of the
physical parameters of the particles $T_{\mathrm{1}}$ and $B_{c}^{\ast -}$
reads
\begin{eqnarray}
&&\Pi _{\mu \delta \nu }^{1\mathrm{Phys}}(p,p^{\prime })=\frac{%
f_{2}(q^{2})\Lambda _{1}f_{B_{c}^{\ast }}^{2}m_{B_{c}^{\ast }}^{2}}{%
(p^{2}-m_{1}^{2})(p^{\prime 2}-m_{B_{c}^{\ast }}^{2})(q^{2}-m_{B_{c}^{\ast
}}^{2})}  \notag \\
&&\times \left[ \epsilon _{\alpha \delta \mu \nu }(p_{\alpha }^{\prime
}+p_{\alpha })+\frac{2p_{\alpha }p_{\beta }^{\prime }\epsilon _{\alpha \beta
\mu \nu }(p_{\delta }-p_{\delta }^{\prime })}{m_{B_{c}^{\ast }}^{2}}\right.
\notag \\
&&\left. +\frac{p_{\alpha }p_{\beta }^{\prime }p_{\mu }^{\prime }\epsilon
_{\alpha \beta \delta \nu }}{m_{B_{c}^{\ast }}^{2}}+\frac{p_{\alpha
}p_{\beta }^{\prime }p_{\nu }\epsilon _{\alpha \beta \delta \mu }}{m_{1}^{2}}%
\right] .  \label{eq:PhysSide2a}
\end{eqnarray}%
The same correlation function obtained using the quark propagators has the
following form
\begin{eqnarray}
&&\Pi _{\mu \delta \nu }^{1\mathrm{OPE}}(p,p^{\prime })=2i^{2}\int
d^{4}xd^{4}ye^{ip^{\prime }y}e^{-ipx}\left\{ \mathrm{Tr}\left[ \gamma
_{\delta }S_{b}^{ia}(-x)\right. \right.  \notag \\
&&\left. \times \gamma _{5}\sigma _{\nu \theta }\widetilde{S}%
_{b}^{jb}(y-x)\gamma _{\mu }\widetilde{S}_{c}^{aj}(x-y)\gamma ^{\theta
}S_{c}^{bi}(x)\right]  \notag \\
&&-\mathrm{Tr}\left[ \gamma _{\delta }S_{b}^{ia}(-x)\gamma _{5}\sigma _{\nu
\theta }\widetilde{S}_{b}^{jb}(y-x)\gamma _{\mu }\widetilde{S}%
_{c}^{bj}(x-y)\right.  \notag \\
&&\left. \left. \times \gamma ^{\theta }S_{c}^{ai}(x)\right] \right\} .
\label{eq:QCDside2a}
\end{eqnarray}%
Having used amplitudes corresponding to the structures $\sim \epsilon
_{\alpha \delta \mu \nu }p_{\alpha }$ both in the physical and QCD
expressions for the correlation function, one can derive SR for the form
factor $f_{2}(q^{2}) $.

At the mass shell $q^{2}=m_{B_{c}^{\ast }}^{2}$ of the $B_{c}^{\ast -}$
meson, the strong coupling $f_{2}$ is equal to
\begin{equation}
f_{2}\equiv \mathcal{F}_{2}(-m_{B_{c}^{\ast }}^{2},m_{1}^{2})=(3.7\pm
0.4)\times 10^{-1}\ ,
\end{equation}%
which is found by means of the function $\mathcal{F}_{2}$ with the
parameters $\mathcal{F}_{2}^{0}=1.1$, $c_{2}^{1}=3.53$, and $%
c_{2}^{2}=-3.58. $

The width of the decay $T_{\mathrm{1}}\rightarrow B_{c}^{\ast -}B_{c}^{\ast
-}$ is determined by the formula
\begin{equation}
\Gamma \left[ T_{\mathrm{1}}\rightarrow B_{c}^{\ast -}B_{c}^{\ast -}\right]
=f_{2}^{2}\frac{\lambda _{2}}{48\pi }\left( 5+5\xi +\frac{8}{\xi }\right) ,
\end{equation}%
where $\lambda _{2}=\lambda (m_{1},m_{B_{c}^{\ast }},m_{B_{c}^{\ast }})$ and
$\xi =m_{_{1}}^{2}/m_{B_{c}^{\ast }}^{2}$. Then, for the width of this
process we get
\begin{equation}
\Gamma \left[ T_{\mathrm{1}}\rightarrow B_{c}^{\ast -}B_{c}^{\ast -}\right]
=(11.8\pm 3.0)~\mathrm{MeV}.
\end{equation}%
Using information obtained in this section, we get  the full width of the
tetraquark $T_{\mathrm{1}}$
\begin{equation}
\Gamma (T_{\mathrm{1}})=(44.3\pm 8.8)~\mathrm{MeV}.
\end{equation}%
which characterizes it as a relatively narrow resonance.


\section{Width of the diquark-antidiquark state $T_{\mathrm{2}}$}

\label{sec:AVWidths2}


The mass of the axial-vector exotic meson $T_{\mathrm{2}}$ is large enough
and makes possible its decays to the final states $B_{c}^{-}B_{c}^{\ast -}$,
$B_{c}^{\ast -}B_{c}^{\ast -}$ and $B_{c}^{-}(2S)B_{c}^{\ast -}$. The meson $%
B_{c}^{-}(2S)$ is a radially excited state of the ground-level particle $%
B_{c}^{-}$. The mesons $B_{c}^{-}$ and $B_{c}^{-}(2S)$ are described by the
same interpolating current $J^{B_{c}}$. This means that the physical side of
the sum rule for the strong coupling $F_{2}$ at the vertex $T_{\mathrm{2}%
}B_{c}^{-}(2S)B_{c}^{\ast -}$ necessarily contains a term corresponding to
the vertex $T_{\mathrm{2}}B_{c}^{-}B_{c}^{\ast -}$ with coupling $F_{1}$.
Contributions of these two terms can be separated by choosing an appropriate
$s_{0}$, i.e., by fixing $s_{0}<$ $\widetilde{m}_{B_{c}}^{2}$ with $%
\widetilde{m}_{B_{c}}$ being the mass of the meson $B_{c}^{-}(2S)$, we can
include effects of $F_{2}$ into "higher resonances and continuum states".
Then, it is not difficult to determine the coupling $F_{1}$, and use it as an 
input parameter in the sum rule with $s_{0}>\widetilde{m}_{B_{c}}^{2}$ to
extract $F_{2}$. Below, we follow namely this strategy.


\subsection{$T_{\mathrm{2}}\rightarrow B_{c}^{-}B_{c}^{\ast -}$ and $T_{%
\mathrm{2}}\rightarrow B_{c}^{-}(2S)B_{c}^{\ast -}$}


The correlation function necessary to obtain the form factors $F_{1}(q^{2})$
and $F_{2}(q^{2})$ at the vertices $T_{\mathrm{2}}B_{c}^{-}B_{c}^{\ast -}$
and $T_{\mathrm{2}}B_{c}^{-}(2S)B_{c}^{\ast -}$ is given by the formula%
\begin{eqnarray}
\Pi _{\mu \nu }^{2}(p,p^{\prime }) &=&i^{2}\int d^{4}xd^{4}ye^{ip^{\prime
}y}e^{-ipx}\langle 0|\mathcal{T}\{J^{B_{c}}(y)  \notag \\
&&\times J_{\mu }^{B_{c}^{\ast }}(0)J_{\nu }^{\dag }(x)\}|0\rangle ,
\label{eq:CF8}
\end{eqnarray}%
The physical side of SRs for $F_{1}(q^{2})$ and $F_{2}(q^{2})$ is determined
by the expression
\begin{eqnarray}
&&\Pi _{\mu \nu }^{2\mathrm{Phys}}(p,p^{\prime })=\frac{\Lambda
_{2}f_{B_{c}^{\ast }}m_{B_{c}^{\ast }}}{(m_{b}+m_{c})\left(
p^{2}-m_{2}^{2}\right) \left( q^{2}-m_{B_{c}^{\ast }}^{2}\right) }  \notag \\
&&\times \left[ F_{1}(q^{2})\frac{f_{B_{c}}m_{B_{c}}^{2}}{(p^{\prime
2}-m_{B_{c}}^{2})}\left( \frac{m_{2}^{2}-m_{B_{c}}^{2}+q^{2}}{2}g_{\mu \nu
}-p_{\mu }q_{\nu }\right) \right.  \notag \\
&&\left. +F_{2}(q^{2})\frac{\widetilde{f}_{B_{c}}\widetilde{m}_{B_{c}}^{2}}{%
(p^{\prime 2}-\widetilde{m}_{B_{c}}^{2})}\left( \frac{m_{2}^{2}-\widetilde{m}%
_{B_{c}}^{2}+q^{2}}{2}g_{\mu \nu }-p_{\mu }q_{\nu }\right) \right]  \notag \\
&&+\cdots .  \label{eq:PhysSide3}
\end{eqnarray}%
Here, $\widetilde{m}_{B_{c}}=(6871.2\pm 1.0)~\mathrm{MeV}$ and $\widetilde{f}%
_{B_{c}}=(420\pm 20)~\mathrm{MeV}$ are parameters of the meson $%
B_{c}^{-}(2S) $ borrowed from Refs.\ \cite{PDG:2022,Aliev:2019wcm},
respectively.

The QCD side of the SRs is:%
\begin{eqnarray}
&&\Pi _{\mu \nu }^{2\mathrm{OPE}}(p,p^{\prime })=2i\int
d^{4}xd^{4}ye^{ip^{\prime }y}e^{-ipx}\left\{ \mathrm{Tr}\left[ \gamma _{\mu
}S_{b}^{ja}(-x)\right. \right.  \notag \\
&&\left. \times \gamma _{5}\gamma _{\nu }\widetilde{S}_{b}^{ib}(y-x)\gamma
_{5}\widetilde{S}_{c}^{ai}(x-y)S_{c}^{bj}(x)\right]  \notag \\
&&\left. -\mathrm{Tr}\left[ \gamma _{\mu }S_{b}^{ja}(-x)\gamma _{5}\gamma
_{\nu }\widetilde{S}_{b}^{ib}(y-x)\gamma _{5}\widetilde{S}%
_{c}^{bi}(x-y)S_{c}^{aj}(x)\right] \right\} .  \notag \\
&&  \label{eq:CF8A}
\end{eqnarray}%
The functions $\Pi _{\mu \nu }^{2\mathrm{Phys}}(p,p^{\prime })$ and $\Pi
_{\mu \nu }^{2\mathrm{OPE}}(p,p^{\prime })$ have two Lorentz structures
proportional to $g_{\mu \nu }$ and $p_{\mu }q_{\nu }$. In what follows, we
employ the amplitudes $\sim g_{\mu \nu }$ to find SR for the form factors $%
F_{1}(q^{2})$ and $F_{2}(q^{2})$. We equate, as usual, the invariant
amplitudes $\Pi _{2}^{\mathrm{Phys}}(p^{2},p^{\prime 2},q^{2})$ and $\Pi
_{2}^{\mathrm{OPE}}(p^{2},p^{\prime 2},q^{2})$ corresponding to these
structures and find an expression which contains two unknown functions $%
F_{1}(q^{2})$ and $F_{2}(q^{2})$.

\begin{figure}[h]
\includegraphics[width=8.5cm]{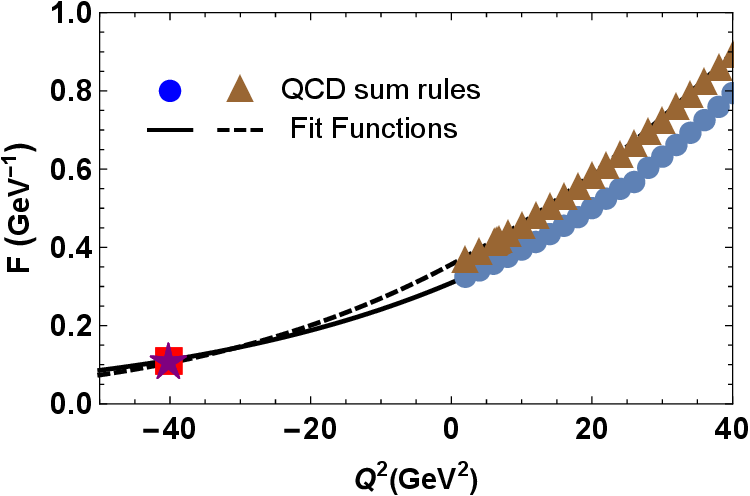}
\caption{QCD data and extrapolating functions $\mathcal{G}_{1}(Q^{2})$
(solid line) and $\mathcal{G}_{2}(Q^{2})$ (dashed line). The star and square
show the points $Q^{2}=-m_{B_{c}^{\ast }}^{2}$ where the couplings $F_{1}$
and $F_{2}$ have been computed. }
\label{fig:Fit1}
\end{figure}

We divide determination of the form factors $F_{1}(q^{2})$ and $F_{2}(q^{2})$
by means of a SR equality into two stages. As the first step, we consider $%
F_{1}(q^{2})$, and employ a result obtained for this form factor at the
second stage to calculate $F_{2}(q^{2})$. In both stages of analysis, we use
for $(M_{1}^{2},s_{0})$ the parameters of Eq.\ (\ref{eq:Wind2}). These two
phases are distinguished by the regions chosen for the parameters $%
(M_{2}^{2},s_{0}^{\prime })$. At first, we limit $s_{0}^{\prime }$ by the
mass of the meson $B_{c}^{-}(2S)$ and fix $s_{0}^{\prime }<\widetilde{m}%
_{B_{c}}^{2}$. This allows us to treat the contribution of the vertex $T_{%
\mathrm{2}}B_{c}^{-}B_{c}^{\ast -}(2S)$ as "a continuum effect" and explore
only the first component in Eq.\ (\ref{eq:PhysSide3}). The parameters $%
(M_{2}^{2},s_{0}^{\prime })$ are determined by the expression
\begin{equation}
M_{2}^{2}\in \lbrack 6.5,7.5]~\mathrm{GeV}^{2}\text{, }s_{0}^{\prime }\in
\lbrack 45,47]~\mathrm{GeV}^{2}.
\end{equation}%
The form factor $F_{1}(q^{2})$ can be modeled by the fit function $\mathcal{G%
}_{1}(Q^{2},m_{2}^{2})$ with parameters $\mathcal{G}_{1}^{0}=0.31\ \mathrm{%
GeV}^{-1}$, $g_{1}^{1}=4.39$ and $g_{1}^{2}=-0.74$ (see, Fig.\ \ref{fig:Fit1}%
). Let us note that the functions $\mathcal{G}_{j}(Q^{2},m_{2}^{2})$ have
the analytic form of Eq.\ (\ref{eq:FitF}) with substitutions $%
m_{1}^{2}\rightarrow m_{2}^{2}$ and $\mathcal{F}_{j}^{0}$, $%
c_{j}^{1(2)}\rightarrow \mathcal{G}_{j}^{0}$, $g_{j}^{1(2)}$, respectively.

The strong coupling $F_{1}$ computed at the mass shell $q^{2}=m_{B_{c}^{\ast
}}^{2}$ of the $B_{c}^{\ast -}$ meson is equal to
\begin{equation}
F_{1}\equiv \mathcal{G}_{1}(-m_{B_{c}^{\ast }}^{2},m_{2}^{2})=(1.1\pm
0.1)\times 10^{-1}\ \mathrm{GeV}^{-1}.
\end{equation}%
At the next step, we choose
\begin{equation}
M_{2}^{2}\in \lbrack 6.5,7.5]~\mathrm{GeV}^{2}\text{, }s_{0}^{\ast \prime
}\in \lbrack 48,50]~\mathrm{GeV}^{2},  \label{eq:Wind3A}
\end{equation}%
and utilize $F_{1}(q^{2})$ as an input to extract the form factor $%
F_{2}(q^{2})$, where $s_{0}^{\ast \prime }$ is limited by the mass of the
meson $B_{c}(3S)$, i.e., $m^{2}[B_{c}(3S)]=(7.272)^{2}~\mathrm{GeV}^{2}$
\cite{Godfrey:2004ya}. Analysis carried out in the context of this approach
leads to the result
\begin{equation}
F_{2}\equiv \mathcal{G}_{2}(-m_{B_{c}^{\ast }}^{2},m_{2}^{2})=(1.0\pm
0.1)\times 10^{-1}\ \mathrm{GeV}^{-1},
\end{equation}%
where $\mathcal{G}_{2}(-m_{B_{c}^{\ast }}^{2},m_{2}^{2})$ is determined by
the parameters $\mathcal{G}_{2}^{0}=0.36\ \mathrm{GeV}^{-1}$, $%
g_{2}^{1}=4.81 $ and $g_{2}^{2}=-3.01$. Relevant information is shown
graphically in Fig.\ \ref{fig:Fit1}.

The width of the decays $T_{\mathrm{2}}\rightarrow B_{c}^{-}B_{c}^{\ast -}$
and $T_{\mathrm{2}}\rightarrow B_{c}^{-}(2S)B_{c}^{\ast -}$, after necessary
refinements, can be computed using Eq.\ (\ref{eq:PDw2}):

\begin{eqnarray}
&&\Gamma \left[ T_{\mathrm{2}}\rightarrow B_{c}^{-}B_{c}^{\ast -}\right]
=(48.1\pm 12.2)~\mathrm{MeV},  \notag \\
&&\Gamma \left[ T_{\mathrm{2}}\rightarrow B_{c}^{-}(2S)B_{c}^{\ast -}\right]
=(19.0\pm 4.8)~\mathrm{MeV}.
\end{eqnarray}

\subsection{$T_{\mathrm{2}}\rightarrow B_{c}^{\ast -}B_{c}^{\ast -}$}


This process is investigated as the decay $T_{\mathrm{1}}\rightarrow
B_{c}^{\ast -}B_{c}^{\ast -}$ considered in the previous section,
differences being in the mass $m_{2}$, correlation function $\Pi _{\mu
\delta \nu }^{2\mathrm{OPE}}(p,p^{\prime })$ and strong coupling $F_{3}$ at
the vertex $T_{\mathrm{2}}B_{c}^{\ast -}B_{c}^{\ast -}$.

The physical side of SR for the form factor $F_{3}(q^{2})$, after some
replacements, is determined by the expression Eq.\ (\ref{eq:PhysSide2a}).
But the function $\Pi _{\mu \delta \nu }^{2\mathrm{OPE}}(p,p^{\prime })$ due
to the current $J_{\mu }^{2}(x)$ takes the following form
\begin{eqnarray}
&&\Pi _{\mu \delta \nu }^{2\mathrm{OPE}}(p,p^{\prime })=2i^{2}\int
d^{4}xd^{4}ye^{ip^{\prime }y}e^{-ipx}\left\{ \mathrm{Tr}\left[ \gamma
_{\delta }S_{b}^{ia}(-x)\right. \right.  \notag \\
&&\left. \times \gamma _{5}\gamma _{\nu }\widetilde{S}_{b}^{jb}(y-x)\gamma
_{\mu }\widetilde{S}_{c}^{aj}(x-y)S_{c}^{bi}(x)\right]  \notag \\
&&\left. +\mathrm{Tr}\left[ \gamma _{\delta }S_{b}^{ia}(-x)\gamma _{5}\gamma
_{\nu }\widetilde{S}_{b}^{jb}(y-x)\gamma _{\mu }\widetilde{S}%
_{c}^{bj}(x-y)S_{c}^{ai}(x)\right] \right\} .  \notag \\
&&
\end{eqnarray}

Remaining manipulations are standard ones, therefore below we provide the
final results obtained for this mode. The coupling $F_{3}$ is computed at
the mass shell of the $B_{c}^{\ast -}$ meson by means of the extrapolating
function $\mathcal{G}_{3}(-m_{B_{c}^{\ast }}^{2},m_{2}^{2})$ with parameters
$\mathcal{G}_{3}^{0}=11.79$, $g_{3}^{1}=13.67$ and $g_{3}^{2}=-20.84$. It is
equal to
\begin{equation}
F_{3}\equiv \mathcal{G}_{3}(-m_{B_{c}^{\ast }}^{2},m_{2}^{2})=(1.9\pm
0.2)\times 10^{-1}.
\end{equation}%
The partial width of the decay $T_{\mathrm{2}}\rightarrow B_{c}^{\ast
-}B_{c}^{\ast -}$ is
\begin{equation}
\Gamma \left[ T_{\mathrm{2}}\rightarrow B_{c}^{\ast -}B_{c}^{\ast -}\right]
=(15.4\pm 3.9)~\mathrm{MeV}.
\end{equation}%
Then, one can easily estimate the full width of the tetraquark $T_{\mathrm{2}%
}$
\begin{equation}
\Gamma (T_{\mathrm{2}})=(82.5\pm 13.7)~\mathrm{MeV}.
\end{equation}


\section{Concluding notes}

\label{sec:Conc}


We have investigated the axial-vector tetraquarks $bb\overline{c}\overline{c}
$ by modeling them as diquark-antidiquark states composed of a diquark $%
b^{T}C\sigma _{\mu \nu }\gamma _{5}b$ and antidiquark $\overline{c}\gamma
^{\nu }C\overline{c}^{T}$ ($T_{\mathrm{1}}$) and a vector diquark and
pseudoscalar antidiquark ($T_{\mathrm{2}}$), respectively. The structures $%
T_{\mathrm{1}}$ and $T_{\mathrm{2}}$ have color triplet $\overline{\mathbf{3}%
_{c}}\otimes \mathbf{3}_{c}{}$ and sextet $\mathbf{6}_{c}\otimes \overline{%
\mathbf{6}}_{c}$ organizations, respectively. Our predictions $%
m_{1}=(12714\pm 86)~\mathrm{MeV}$ and $m_{2}=(13383\pm 92)~\mathrm{MeV}$
prove that these tetraquarks are unstable against the strong decays. In this
aspect, our conclusions are in accord with ones made in Refs.\ \cite%
{Wu:2016vtq,Wang:2019rdo,Liu:2019zuc,Galkin:2023wox}. But, we could not
confirm predictions made in Refs.\ \cite{Wang:2021taf,Mutuk:2022nkw} about
the strong-interaction stable nature some of the axial-vector particles $bb%
\overline{c}\overline{c}$.

The results for the masses of the structures $T_{\mathrm{1}}$ and $T_{%
\mathrm{2}}$ have permitted us to reveal their possible decay modes. The
full width of the exotic mesons $T_{\mathrm{1}}$ and $T_{\mathrm{2}}$ are
evaluated by computing partial widths of the decays $T_{\mathrm{1}%
}\rightarrow B_{c}^{-}B_{c}^{\ast -}$, $B_{c}^{\ast -}B_{c}^{\ast -}$ and $%
T_{\mathrm{2}}\rightarrow B_{c}^{-}B_{c}^{\ast -}$, $B_{c}^{\ast
-}B_{c}^{\ast -}$ and $T_{\mathrm{2}}\rightarrow B_{c}^{-}(2S)B_{c}^{\ast -}$%
, respectively. Predictions for the full widths of the axial-vector
tetraquarks $\Gamma (T_{\mathrm{1}})=(44.3\pm 8.8)~\mathrm{MeV}$ and $\Gamma
(T_{\mathrm{2}})=(82.5\pm 13.7)~\mathrm{MeV}$ mean\ that they may be
interpreted as states with modest widths.

As is seen there are controversial results for the parameters of the $bb%
\overline{c}\overline{c}$/$cc\overline{b}\overline{b}$ tetraquarks with
spin-parities $J^{\mathrm{P}}=0^{+}$, $1^{+}$ and $2^{+}$. Additionally, our
analyses do not encompass all possible axial-vector states which may be
composed using diquarks (antidiquarks) with different quantum numbers. Such
structures may be also studied in the sum rule framework. Because the
four-quark compounds $bb\overline{c}\overline{c}$/$cc\overline{b}\overline{b}
$ did not yet discovered experimentally, it is difficult to make conclusions
about features of such particles. They may be pure $T_{\mathrm{1}}$ or $T_{%
\mathrm{2}}$ states and bear parameters of these structures. Alternatively,
physical resonances may be a superposition of these and other basic states.

In any case, further experimental and theoretical studies of multiquark
mesons $bb\overline{c}\overline{c}$/$cc\overline{b}\overline{b}$ are
required for reliable statements concerning parameters of such tetraquarks.
Our present analysis is a useful step in this direction.

\section*{ACKNOWLEDGEMENTS}

K. Azizi is thankful to Iran National Science Foundation (INSF) for the
partial financial support provided under the elites Grant No. 4025036.

\end{document}